\documentclass[aip,amsmath,amssymb,preprint]{revtex4-2}
\usepackage{color}
\usepackage{graphicx}
\usepackage{amssymb}
\usepackage{amsmath}
\usepackage{todonotes}
\usepackage[hidelinks=true]{hyperref}
\hypersetup{
		colorlinks = true,
		urlcolor = blue,
		linkcolor = blue,
		citecolor = blue
	}
\usepackage[utf8]{inputenc}
\usepackage[T1]{fontenc}
\usepackage{mathptmx}
\usepackage{etoolbox}

\makeatletter
\def\@email#1#2{%
 \endgroup
 \patchcmd{\titleblock@produce}
  {\frontmatter@RRAPformat}
  {\frontmatter@RRAPformat{\produce@RRAP{*#1\href{mailto:#2}{#2}}}\frontmatter@RRAPformat}
  {}{}
}%
\makeatother

\begin{document}

\title{Diffusion-assisted molecular beam epitaxy of CuCrO$_2$ thin films }

%% authors 
\author{Gaurab Rimal}
\email{gr380@physics.rutgers.edu}
\affiliation{Department of Physics \& Astronomy, Rutgers, The State University of New Jersey, Piscataway, New Jersey 08854, USA }

\author{Alessandro R. Mazza}
\affiliation{Materials Science and Technology Division, Oak Ridge National Laboratory, Oak Ridge, Tennessee 37831, USA }
\affiliation{Center for Integrated Nanotechnologies, Los Alamos National Laboratory, Los Alamos, New Mexico, USA}

\author{Matthew Brahlek}
\affiliation{Materials Science and Technology Division, Oak Ridge National Laboratory, Oak Ridge, Tennessee 37831, USA }

\author{Seongshik Oh}
\email{ohsean@physics.rutgers.edu}
\affiliation{Department of Physics \& Astronomy, Rutgers, The State University of New Jersey, Piscataway, New Jersey 08854, USA }
\affiliation{Center for Quantum Materials Synthesis, Rutgers, The State University of New Jersey, Piscataway, New Jersey 08854, USA }

%%%% macros
\newcommand{\red}[1]{\textcolor{red}{#1}}
\newcommand{\blue}[1]{\textcolor{blue}{#1}}
\newcommand{\green}[1]{\textcolor{green}{#1}}

\begin{abstract}
Using molecular beam epitaxy (MBE) to grow multi-elemental oxides (MEO) is generally challenging, partly due to difficulty in stoichiometry control. Occasionally, if one of the elements is volatile at the growth temperature, stoichiometry control can be greatly simplified using adsorption-controlled growth mode. Otherwise, stoichiometry control remains one of the main hurdles to achieving high quality MEO film growths. Here, we report another kind of self-limited growth mode, dubbed diffusion-assisted epitaxy, in which excess species diffuses into the substrate and leads to the desired stoichiometry, in a manner similar to the conventional adsorption-controlled epitaxy. Specifically, we demonstrate that using diffusion-assisted epitaxy, high-quality epitaxial CuCrO$_2$ films can be grown over a wide growth window without precise flux control using MBE.

\end{abstract}

\maketitle

Transition metal oxides (TMO) provide a rich playground for fundamental and applied research, in which competitions between various degrees of freedom develop novel phases \cite{Hwang2012}. Epitaxial thin films provide a great way to engineer the properties of TMO through the control of thickness, strain and heterostructures. Although single crystals of TMO can be grown readily in bulk form, there are inherent challenges associated with the growth of epitaxial films. Following significant progress since the discovery of high-T$_C$ cuprate superconductors, many binary, ternary and higher order TMO can now be grown epitaxially. Modern epitaxy methods, such as molecular beam epitaxy (MBE), pulsed laser deposition/epitaxy (PLD/PLE) and reactive sputtering have been adapted to grow high-quality films of TMO. However, growing TMO can be challenging for MBE when materials contain species that are difficult to oxidize: in such a case, strong oxidants such as ozone \cite{Vasko1995,Bozovic2002,Krockenberger2008,Schlom2015,Nair2018} or oxygen plasma \cite{Chambers2002} must be used. Similar issues are also encountered in other growth methods such as PLD and sputtering, so many materials are grown under or annealed in a high oxygen pressure environment \cite{Boris2011,Chakhalian2006,Frano2013,Gao1991}. %If these challenges can be circumvented, films grown using OMBE usually exhibit properties which are the best among all thin film methods \cite{Schlom2015}. 

 %%% Oxide MBE review \cite{Schlom2015,Nunn2021}
 %%% PLD review \cite{Christen2008} 

Another big challenge in multi-elemental TMO (ME-TMO) by MBE is stoichiometry control. One of the main factors that made MBE so successful for creating the highest quality compound semiconductors is the development of adsorption-controlled or self-limited growth mode, in which excess volatile species re-effuse from the surface, naturally leading to perfect stoichiometry \cite{Henini2012,Pfeiffer2003,Falson2022}. For binary oxides, adsorption-controlled growth mode is applicable with oxygen working as the self-limited species. However, for multi-elemental oxides, such a self-limited growth mode is generally not possible unless one of the elements is a highly volatile species such as Pb and Bi \cite{Hellman1990,Theis1998,Theis1998b}. Another MBE variant called hybrid MBE \cite{Jalan2009,Brahlek2018}, which uses volatile metal-organic chemical flux, is also used to grow some stoichiometric ME-TMOs. However, for most ME-TMOs, stoichiometry control remains one of the most challenging tasks for achieving high quality films using MBE. 

In typical MBE growth conditions, bulk diffusion of metal elements is negligible compared to surface diffusion. This is because the growth temperature used in MBE is typically lower than that required for bulk crystal growth. In fact, this suppressed bulk diffusion is frequently utilized to form thermodynamically metastable artificially layered heterostructures. However, if bulk diffusion of one of the elements is significant during the film growth, particularly into the substrate, the substrate can serve like the vacuum in the adsorption controlled growth. If such a "reverse adsorption control" is possible, we should be able to achieve similar self-limited growth mode and grow high quality stoichiometric films without precise flux control of each element. Here, we show that such a diffusion-assisted self-limited growth mode is indeed possible when CuCrO$_2$ films are grown on sapphire substrate. It turns out that excess copper can diffuse readily not only through the bulk of CuCrO$_2$ but also through the sapphire substrate, leading to a phase-pure epitaxial film over a wide growth window.

CuCrO$_2$ belongs to the delafossite family, a unique class of layered triangular TMO with a general chemical formula of ABO$_2$ in which A and B are monovalent and trivalent transition metal species, respectively. The crystal has layered A and BO$_2$ molecular blocks and exhibits properties that are strongly dependent on the A and B species \cite{Marquardt2006,Mackenzie2017}. For example, behaviors such as hydrodynamic and anisotropic transport \cite{Hicks2012,Moll2016,Mackenzie2017}, spin-split surface \cite{Mazzola2017} and metamagnetic transitions \cite{Coldea2014} have been observed in high-quality crystals \cite{Sunko2020}. CuCrO$_2$ is an interesting material on its own. It is a semiconductor with a bandgap of about 3 eV \cite{Scanlon2011} and is a candidate for a p-type transparent conductor \cite{Chiba2017,Chiba2018,Tripathi2017}. It is also an antiferromagnet with helical ordering and exhibits multiferroicity below T$_N \sim $ 24 K \cite{Kimura2006,Seki2008,Poienar2010,Frontzek2011}. Furthermore, CuCrO$_2$ can be used as a buffer layer for the growth of other delafossites such as PdCrO$_2$ \cite{Ok2020}.

Growth of delafossites by MBE is especially tricky. First, since Cu, Pd, Pt and Ag cannot be fully oxidized using molecular oxygen, activated oxygen sources such as atomic oxygen or ozone are necessary \cite{Shin2012,Brahlek2019,Sun2019}. Furthermore, due to the presence of competing phases, the growth window is small and relatively low growth temperatures are needed to avoid the secondary phases, resulting in low sample quality. None of these elements are volatile at typical growth temperatures so adsorption-controlled growth is generally not possible, making it also difficult to achieve the right stoichiometry. However, the fact that many of these elements also diffuse readily into substrates \cite{Bracht2004,Doremus2006} provides the alternative scenario of \textit{diffusion-assisted growth} to be envisioned.

%\section*{General experimental conditions} 
For this study, we grew CuCrO$_2$ films on c-axis oriented Al$_2$O$_3$ substrates using MBE. Growth temperature is varied between 600 $^\circ$C and 1150 $^\circ$C while the background oxygen gas pressure is fixed at 4$\times 10^{-6}$ Torr. Oxygen plasma was generated via a standard RF plasma generator at 13.6 MHz at a power of 450 W. The gas port is located about 20 cm from the surface of the sample. The growth was monitored in-situ using reflection high-energy electron diffraction (RHEED). We found that single phase films can only be obtained using layer-by-layer growth, and co-deposition of Cu and Cr leads to multiphase films. Therefore, we will focus our discussion on films grown in a layer-by-layer fashion via sequential shuttering of Cr and Cu sources. All films are 42 monolayers (ML) thick, corresponding to about 25 nm. Cr flux was fixed at about 2.5$\times 10^{13}$ atoms/cm$^2$s, while Cu fluxes were varied with respect to Cr: flux of each element was measured using quartz crystal microbalance (QCM) and calibrated using Rutherford backscattering spectrometry (RBS). RBS fitting was done using SIMNRA program, and inclusion of roughness and porosity in the fitting did not change the fitting results.

%\section*{R\lowercase{ole of oxidant } }

\begin{figure}[ht]
	\centering
	\includegraphics{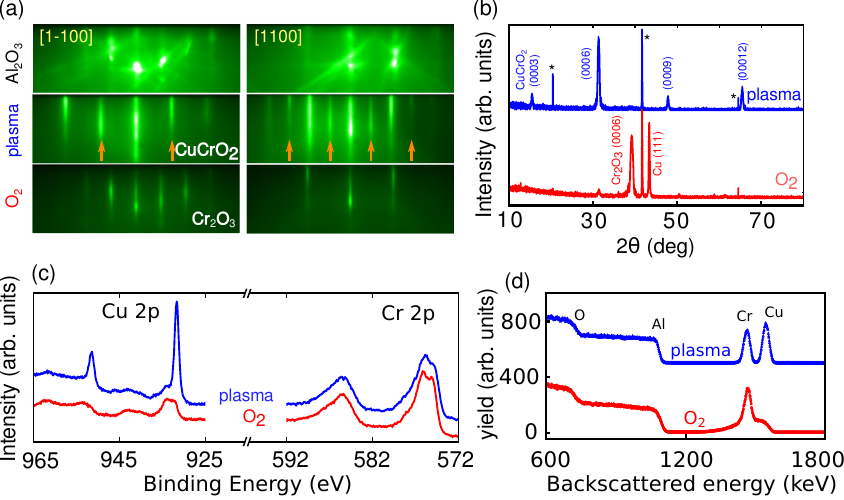}
	\caption{Film growths at 700 $^{\circ}$C using Cu/Cr=1 with plasma and molecular oxygen. (a) RHEED for film grown using plasma shows streaks corresponding to delafossite phase while use of molecular oxygen results in the growth of Cr$_2$O$_3$ plus Cu. The streaks marked by arrows are second order reconstructions. (b) X-ray diffraction comparing films grown using the two oxygen sources. Substrate contributions are marked by *. (c) Cu 2p and Cr 2p XPS spectra. (d) RBS spectra taken at incident energy of 2.0 MeV comparing the films grown using O$_2$ and plasma. In the case of O$_2$ grown film, the extended, asymmetric tail results from backscattering of He$^{++}$ from Cu diffused into the substrate. } 
	\label{fig1}
\end{figure}
%%%%%%%%%%%%

Techniques such as plasma assisted MBE \cite{Shin2012} and PLD \cite{OSullivan2010,Ok2020b} were previously employed in the growth of CuCrO$_2$ films. However, the MBE-grown films showed polycrystalline grains, and no detailed study of the growth and nature of the films at higher temperatures was undertaken. We first study the role of oxidant and Cu diffusion on the growth of CuCrO$_2$. 

Figure \ref{fig1} shows the importance of activated source. At the growth temperature of 700 $^\circ$C with Cu/Cr=1, while CuCrO$_2$ can be readily grown in plasma, Cr$_2$O$_3$ phase dominates when grown in O$_2$. RHEED and x-ray diffraction (XRD) in Figure \ref{fig1}(a,b) show the formation of the proper delafossite phase with plasma, with lattice constants a = 2.97 \textrm{\AA} and c = 17.08 \textrm{\AA}. But the growth under molecular oxygen is different. In such a case, RHEED shows that the lattice constant slowly relaxes to the value of 4.94 \textrm{\AA}, suggesting the growth of Cr$_2$O$_3$ instead. XRD, as shown in Figure \ref{fig1}(b), verifies that Cr$_2$O$_3$ is the dominant phase in the film. X-ray photoelectron spectroscopy (XPS) shown in Fig. \ref{fig1}(c) also confirms that the two films are different. Since Cr in both Cr$_2$O$_3$ and CuCrO$_2$ is trivalent, the Cr spectra look similar. However, the Cu spectra are different. While a clear Cu$^{1+}$ state can be observed for the film prepared using plasma \cite{Poulston1996,Shin2012}, the other film shows broadened and reduced peaks, indicating low amount of Cu on the surface with multiple oxidation states. 

It is intriguing that Cr$_2$O$_3$ phase forms with molecular oxygen. XRD shows the presence of metallic Cu in the bulk, but RHEED shows that the surface has long-range coherence and high quality. It is likely that Cu may have diffused deeper into the film or the substrate. We verify Cu diffusion using RBS, as shown in Figure \ref{fig1}(d). In RBS, He$^{++}$ backscatters from heavier elements, and the yield (intensity) depends on the incident energy, the geometry of the setup, the thickness profile of the film and the species in the film that leads to scattering \cite{Feldman1986}. This can be parametrized as $Y = N_s \, \sigma(\theta) \, Q_0 \, \Omega \, $, where $\sigma$ is the scattering cross section for a given scattering angle $\theta$, $\Omega$ is the solid angle for the detector, $Q$ is the incident He$^{++}$ flux, and $N_s$ is the number of scattering atoms. It should be noted that the backscattering profile depends on the overall thickness (or total number of target atoms) that lead to scattering, and a symmetric profile can be expected in the case of thin and uniform films. The film grown under plasma shows two symmetric peaks corresponding to backscattering off Cr and Cu. However, the film grown in molecular oxygen shows an asymmetric and skewed profile for Cu which overlaps with the contribution from Cr. There is a long tail towards lower energies as a result of scattering from lower concentration of Cu atoms from deeper parts of the substrate. Cu is known to diffuse readily into Al$_2$O$_3$ \cite{Moya1993,Doremus2006} and other substrates such as Si and Ge \cite{Bracht2004}, and diffusion is enhanced in oxygen \cite{Gai1990}. Assuming steady diffusion, we can reasonably fit the RBS spectrum using a diffusion profile through the substrate (see supplemental Figure S1)\cite{supplementary} under a constant diffusion constant of the order $10^{-14}$ cm$^2$/s at 700 $^\circ$C.

%%%%%%%% Figure 2 for 
\begin{figure}[ht]
	\centering
	\includegraphics{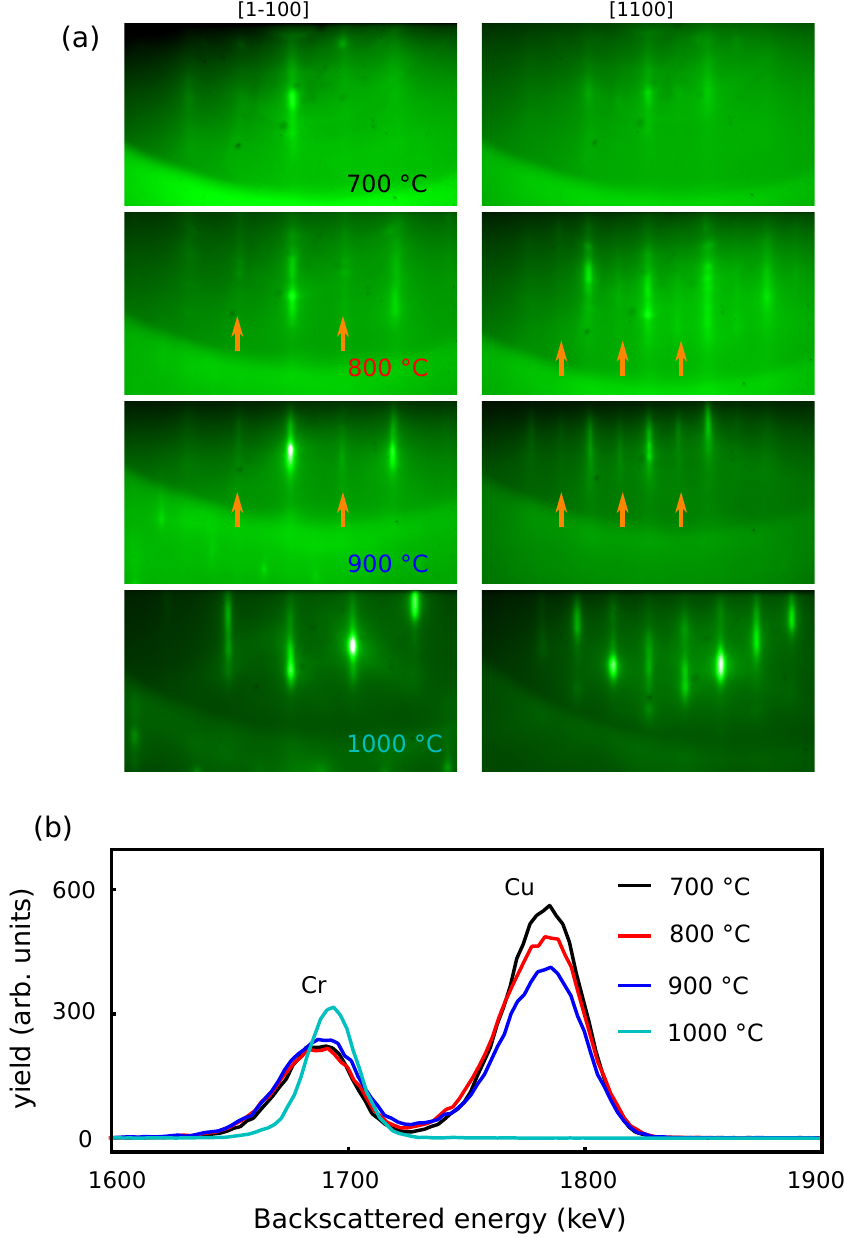}
	\caption{Comparison of CuCrO$_2$ grown in plasma at different temperatures using Cu/Cr=1.5. (a) RHEED shows that as growth temperature increases up to 900 $^\circ$C, the diffraction corresponding to the CuCrO$_2$ phase becomes more coherent. However, at 1000 $^\circ$C, only Cr$_2$O$_3$ phase is observable. Surface reconstruction streaks in CuCrO$_2$ are marked by arrows. (b) RBS at 2.3 MeV for the three films showing Cu diffusion at different growth temperatures. Absence of Cu peak at 1000 $^\circ$C is consistent with the RHEED pattern shown in (a). }
	\label{fig2}
\end{figure}

One key observation for our growth is that when excess Cu is used during growth, even under plasma, an asymmetric profile for Cu appears in the RBS spectra. To understand this behavior, we grow samples using excess Cu, and test the role of elevated growth temperatures, which leads to higher diffusion. In Figure \ref{fig2}, we investigate the role of growth temperature on the film properties and diffusion. RHEED in Figure \ref{fig2}(a) shows how the film quality is affected by excess Cu (1.5 times stoichiometric value) at different temperatures. At the lowest growth temperature of 700 $^\circ$C, the surface has almost ring-like features along with hazy streaks that correspond to CuCrO$_2$, showing lower quality with possibly polycrystalline or multiple phases. At the growth temperature of 800 $^\circ$C, the CuCrO$_2$ streaks become more pronounced, and at 900 $^\circ$C, they become even better, without any hint of defective RHEED patterns. However, when grown at a temperature of 1000 $^\circ$C, the RHEED pattern completely changes to that of Cr$_2$O$_3$ without any hint of CuCrO$_2$.  

RBS, shown in Figure \ref{fig2}(b), helps explain what occurs at different growth temperatures. Although not as severe as the film grown in O$_2$, asymmetric Cu peaks indicate diffusion of Cu into the substrate. When the growth temperature is 700 $^\circ$C, the asymmetric Cu and Cr peaks have the highest area. At 800 $^\circ$C and 900 $^\circ$C, the profiles are similar, but area under the Cu peak decreases while that of Cr slightly increases, indicating that Cu diffused deeper into the substrate or re-evaporated from the surface. However, at 1000 $^\circ$C, no signature of Cu is observed, and Cr has a more symmetric profile, suggesting that at this growth temperature, Cu either did not stick to the film or completely diffused deeply into the substrate. Such a condition may be more appropriate for adsorption-controlled growth with excessive Cu flux, and this possibility is discussed in the supplemental Figure S2 \cite{supplementary}. We also emphasize that the sticking coefficient of Cu may decrease with increasing temperature, which could result in the decrease in the Cu peak intensity/area with increasing temperature.

%\section*{Role of Cu flux and growth temperature} 

%%%%%%%% Figure 3 for 
\begin{figure}[ht]
	\centering
	\includegraphics{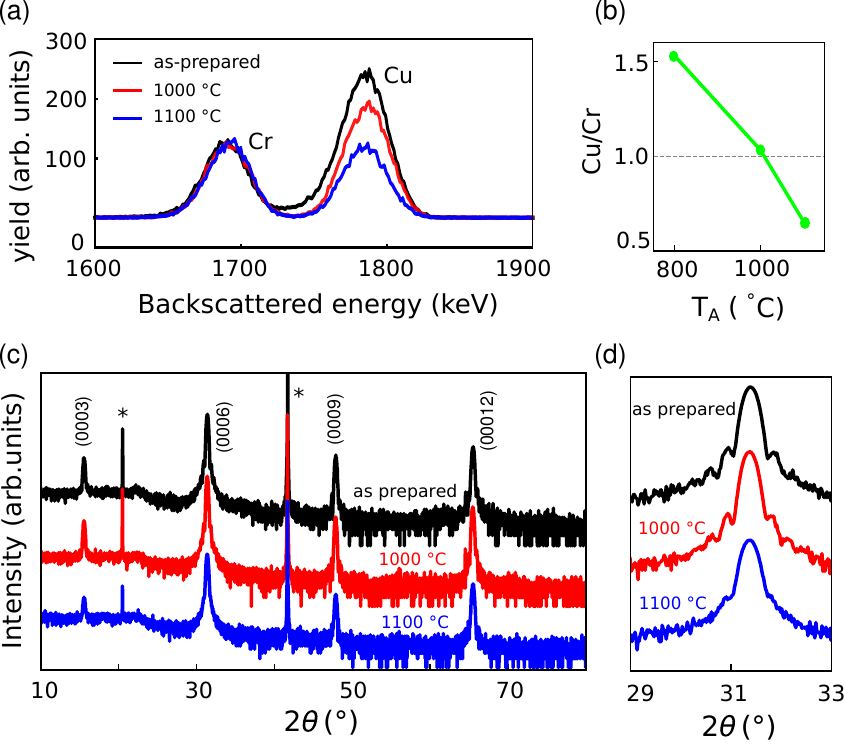}
	\caption{CuCrO$_2$ film grown in plasma at 800 $^\circ$C with Cu/Cr=1.5 and annealed in-situ at the indicated temperatures. (a) RBS at 2.3 MeV. With increasing anneal temperatures, Cu profile becomes more symmetric, suggesting that the excess Cu staying near the top of the substrate diffuses away deeply into the substrate or re-evaporates. (b) Cation stoichiometry at different anneal temperatures determined from the RBS. (c) Annealing-temperature-dependent XRD spectra. Almost identical XRD spectra suggest that the film quality does not change much up to 1100 $^\circ$C  of annealing. The substrate peaks are marked by *. (d) Zoomed-in view of the (0006) peaks. The quality of Laue oscillations around the peak shows that the film quality actually worsens after annealing at 1100 $^\circ$C. }
	\label{fig3}
\end{figure}
%%%%%%%%%%%%

The above scenario suggests that bulk diffusion could be used as an effective parameter to facilitate the growth of epitaxial films. Since monovalent Cu diffuses readily into many substrates \cite{Doremus2006}, it may be possible to have a diffusion-assisted growth regime with excess Cu, in which excess Cu diffuses into the substrate leaving fully stoichiometric film behind. In Figure \ref{fig3}(a), we show RBS for samples that were grown at 800 $^\circ$C using excess Cu (Cu/Cr=1.5), and subsequently annealed to higher temperatures in plasma. For the as-prepared sample, diffusion leads to asymmetry in Cu peak, and XRD shows no secondary phases for this film, while Laue oscillations reveal high quality surface and interface. After annealing to higher temperatures, the Cu peak becomes more symmetric with substantially reduced tail, suggesting that the extra Cu staying near the top of the substrate has diffused deeply into the substrate and/or re-evaporated into the vacuum. Figure \ref{fig3}(b) shows that the cation stoichiometry derived from RBS approaches unity at T$_A$ = 1000 $^\circ$C, while it falls below unity at the highest anneal temperature of 1100 $^\circ$C, likely as a result of film degradation. 

The high quality of these epitaxial films, even when using excess Cu, is evidenced by XRD in Figure \ref{fig3}(c). There is no change in the c-axis lattice constant up to 1100 $^\circ$C of annealing temperature, suggesting that the excess Cu is lost, likely due to diffusion or re-evaporation, resulting in a stoichiometric CuCrO$_2$ film. However, as shown in Figure \ref{fig3}(d), annealing at 1100 $^\circ$C causes the Laue oscillations to dampen, suggesting the degradation of the interface quality. At an even higher annealing temperature of 1150 $^\circ$C, RHEED (not reported here) shows complete loss of surface coherence, as a result of significant degradation of the film quality. These observations suggest that with postgrowth annealing, excess Cu can be diffused or evaporated away up to 1000 $^\circ$C of annealing while improving the quality of the film, but near or higher than 1100 $^\circ$C, the film quality starts to degrade with significant loss of Cu. 

This novel method of MBE growth provides an opportunity to expand the growth window and grow materials by harnessing the diffusion of metal solutes in the bulk subtrate. Many substrates, such as Si, Ge and Al$_2$O$_3$ exhibit diffusion of metal species, as discussed above. By harnessing such a scheme and studying the role of solute diffusion in different substrates, high-quality epitaxial materials may be obtained. Notably, noble metal species such as Cu, Ag, Pt and Au may exhibit such behavior across many substrates. For growth of oxides, substrates such as Si and Ge are prone to self-oxidation and may not be applicable, and other substrates are needed. Studies of the interaction of different types of solutes and substrates may help in identifying similar growth schemes for various types of materials.  

In conclusion, we have demonstrated the use of diffusion-assisted epitaxy by exploiting Cu diffusion into the substrate to stabilize epitaxial CuCrO$_2$ films. Phase-pure films of CuCrO$_2$ can be obtained even though excess Cu is used, mainly because excess Cu diffuses into the substrate. This epitaxy mode can be considered ``reverse adsorption controlled'' in the sense that the excess species diffuse into the substrate instead of (or in addition to) evaporating away into the vacuum, yielding phase-pure epitaxial films. Specifically, with Cu/Cr=1 and plasma oxygen, phase-pure epitaxial CuCrO$_2$ film can be obtained only near 700 $^\circ$C of growth temperature, whereas with Cu/Cr=1.5, the phase-pure film can be obtained at growth temperatures up to 900 $^\circ$C and annealing temperatures up to 1100 $^\circ$C, with excess Cu diffused or evaporated away. In other words, with the diffusion-assisted-epitaxy scheme, the growth window can be substantially expanded without strict flux control. Similar scheme may be applicable to the growth of other multi-elemental oxides containing noble metal species.

\section*{Acknowledgements}
We acknowledge support from National Science Foundation (NSF) Grant No. DMR2004125 and Army Research Office (ARO) Grant No. W911NF2010108. The work at Oak Ridge National Lab was supported by the U. S. Department of Energy (DOE), Office of Science, Basic Energy Sciences (BES), Materials Sciences and Engineering Division. We thank Hussein Hijazi and Ryan Thorpe for RBS measurements. 

\section*{Author Declarations}
The authors have no conflicts to disclose. 

\section*{Data availability}
The raw data used for the figures in the main text and the supplemental material can be obtained from the corresponding author with a reasonable request.

%\bibliographystyle{unsrt}
%\bibliography{main}

\begin{thebibliography}{47}%
\makeatletter
\providecommand \@ifxundefined [1]{%
 \@ifx{#1\undefined}
}%
\providecommand \@ifnum [1]{%
 \ifnum #1\expandafter \@firstoftwo
 \else \expandafter \@secondoftwo
 \fi
}%
\providecommand \@ifx [1]{%
 \ifx #1\expandafter \@firstoftwo
 \else \expandafter \@secondoftwo
 \fi
}%
\providecommand \natexlab [1]{#1}%
\providecommand \enquote  [1]{``#1''}%
\providecommand \bibnamefont  [1]{#1}%
\providecommand \bibfnamefont [1]{#1}%
\providecommand \citenamefont [1]{#1}%
\providecommand \href@noop [0]{\@secondoftwo}%
\providecommand \href [0]{\begingroup \@sanitize@url \@href}%
\providecommand \@href[1]{\@@startlink{#1}\@@href}%
\providecommand \@@href[1]{\endgroup#1\@@endlink}%
\providecommand \@sanitize@url [0]{\catcode `\\12\catcode `\$12\catcode
  `\&12\catcode `\#12\catcode `\^12\catcode `\_12\catcode `\%12\relax}%
\providecommand \@@startlink[1]{}%
\providecommand \@@endlink[0]{}%
\providecommand \url  [0]{\begingroup\@sanitize@url \@url }%
\providecommand \@url [1]{\endgroup\@href {#1}{\urlprefix }}%
\providecommand \urlprefix  [0]{URL }%
\providecommand \Eprint [0]{\href }%
\providecommand \doibase [0]{https://doi.org/}%
\providecommand \selectlanguage [0]{\@gobble}%
\providecommand \bibinfo  [0]{\@secondoftwo}%
\providecommand \bibfield  [0]{\@secondoftwo}%
\providecommand \translation [1]{[#1]}%
\providecommand \BibitemOpen [0]{}%
\providecommand \bibitemStop [0]{}%
\providecommand \bibitemNoStop [0]{.\EOS\space}%
\providecommand \EOS [0]{\spacefactor3000\relax}%
\providecommand \BibitemShut  [1]{\csname bibitem#1\endcsname}%
\let\auto@bib@innerbib\@empty
%</preamble>
\bibitem [{\citenamefont {Hwang}\ \emph {et~al.}(2012)\citenamefont {Hwang},
  \citenamefont {Iwasa}, \citenamefont {Kawasaki}, \citenamefont {Keimer},
  \citenamefont {Nagaosa},\ and\ \citenamefont {Tokura}}]{Hwang2012}%
  \BibitemOpen
  \bibfield  {author} {\bibinfo {author} {\bibfnamefont {H.~Y.}\ \bibnamefont
  {Hwang}}, \bibinfo {author} {\bibfnamefont {Y.}~\bibnamefont {Iwasa}},
  \bibinfo {author} {\bibfnamefont {M.}~\bibnamefont {Kawasaki}}, \bibinfo
  {author} {\bibfnamefont {B.}~\bibnamefont {Keimer}}, \bibinfo {author}
  {\bibfnamefont {N.}~\bibnamefont {Nagaosa}},\ and\ \bibinfo {author}
  {\bibfnamefont {Y.}~\bibnamefont {Tokura}},\ }\href
  {https://doi.org/10.1038/nmat3223} {\bibfield  {journal} {\bibinfo  {journal}
  {Nat. Mater.}\ }\textbf {\bibinfo {volume} {11}},\ \bibinfo {pages}
  {103--113} (\bibinfo {year} {2012})}\BibitemShut {NoStop}%
\bibitem [{\citenamefont {Vas'Ko}\ \emph {et~al.}(1995)\citenamefont {Vas'Ko},
  \citenamefont {Nordman}, \citenamefont {Kraus}, \citenamefont {Achutharaman},
  \citenamefont {Ruosi},\ and\ \citenamefont {{Goldman And}}}]{Vasko1995}%
  \BibitemOpen
  \bibfield  {author} {\bibinfo {author} {\bibfnamefont {V.~A.}\ \bibnamefont
  {Vas'Ko}}, \bibinfo {author} {\bibfnamefont {C.~A.}\ \bibnamefont {Nordman}},
  \bibinfo {author} {\bibfnamefont {P.~A.}\ \bibnamefont {Kraus}}, \bibinfo
  {author} {\bibfnamefont {V.~S.}\ \bibnamefont {Achutharaman}}, \bibinfo
  {author} {\bibfnamefont {A.~R.}\ \bibnamefont {Ruosi}},\ and\ \bibinfo
  {author} {\bibfnamefont {A.~M.}\ \bibnamefont {{Goldman And}}},\ }\href
  {https://doi.org/10.1063/1.116187} {\bibfield  {journal} {\bibinfo  {journal}
  {Appl. Phys. Lett.}\ }\textbf {\bibinfo {volume} {2571}},\ \bibinfo {pages}
  {2571} (\bibinfo {year} {1995})}\BibitemShut {NoStop}%
\bibitem [{\citenamefont {Bozovic}\ \emph {et~al.}(2002)\citenamefont
  {Bozovic}, \citenamefont {Logvenov}, \citenamefont {Belca}, \citenamefont
  {Narimbetov},\ and\ \citenamefont {Sveklo}}]{Bozovic2002}%
  \BibitemOpen
  \bibfield  {author} {\bibinfo {author} {\bibfnamefont {I.}~\bibnamefont
  {Bozovic}}, \bibinfo {author} {\bibfnamefont {G.}~\bibnamefont {Logvenov}},
  \bibinfo {author} {\bibfnamefont {I.}~\bibnamefont {Belca}}, \bibinfo
  {author} {\bibfnamefont {B.}~\bibnamefont {Narimbetov}},\ and\ \bibinfo
  {author} {\bibfnamefont {I.}~\bibnamefont {Sveklo}},\ }\href
  {https://doi.org/10.1103/PhysRevLett.89.107001} {\bibfield  {journal}
  {\bibinfo  {journal} {Phys. Rev. Lett.}\ }\textbf {\bibinfo {volume} {89}},\
  \bibinfo {pages} {107001} (\bibinfo {year} {2002})}\BibitemShut {NoStop}%
\bibitem [{\citenamefont {Krockenberger}\ \emph {et~al.}(2008)\citenamefont
  {Krockenberger}, \citenamefont {Kurian}, \citenamefont {Winkler},
  \citenamefont {Tsukada}, \citenamefont {Naito},\ and\ \citenamefont
  {Alff}}]{Krockenberger2008}%
  \BibitemOpen
  \bibfield  {author} {\bibinfo {author} {\bibfnamefont {Y.}~\bibnamefont
  {Krockenberger}}, \bibinfo {author} {\bibfnamefont {J.}~\bibnamefont
  {Kurian}}, \bibinfo {author} {\bibfnamefont {A.}~\bibnamefont {Winkler}},
  \bibinfo {author} {\bibfnamefont {A.}~\bibnamefont {Tsukada}}, \bibinfo
  {author} {\bibfnamefont {M.}~\bibnamefont {Naito}},\ and\ \bibinfo {author}
  {\bibfnamefont {L.}~\bibnamefont {Alff}},\ }\href
  {https://doi.org/10.1103/PhysRevB.77.060505} {\bibfield  {journal} {\bibinfo
  {journal} {Phys. Rev. B}\ }\textbf {\bibinfo {volume} {77}},\ \bibinfo
  {pages} {060505(R)} (\bibinfo {year} {2008})}\BibitemShut {NoStop}%
\bibitem [{\citenamefont {Schlom}(2015)}]{Schlom2015}%
  \BibitemOpen
  \bibfield  {author} {\bibinfo {author} {\bibfnamefont {D.~G.}\ \bibnamefont
  {Schlom}},\ }\href {https://doi.org/10.1063/1.4919763} {\bibfield  {journal}
  {\bibinfo  {journal} {APL Mater.}\ }\textbf {\bibinfo {volume} {3}},\
  \bibinfo {pages} {062403} (\bibinfo {year} {2015})}\BibitemShut {NoStop}%
\bibitem [{\citenamefont {Nair}\ \emph {et~al.}(2018)\citenamefont {Nair},
  \citenamefont {Liu}, \citenamefont {Ruf}, \citenamefont {Schreiber},
  \citenamefont {Shang}, \citenamefont {Baek}, \citenamefont {Goodge},
  \citenamefont {Kourkoutis}, \citenamefont {Liu}, \citenamefont {Shen},\ and\
  \citenamefont {Schlom}}]{Nair2018}%
  \BibitemOpen
  \bibfield  {author} {\bibinfo {author} {\bibfnamefont {H.~P.}\ \bibnamefont
  {Nair}}, \bibinfo {author} {\bibfnamefont {Y.}~\bibnamefont {Liu}}, \bibinfo
  {author} {\bibfnamefont {J.~P.}\ \bibnamefont {Ruf}}, \bibinfo {author}
  {\bibfnamefont {N.~J.}\ \bibnamefont {Schreiber}}, \bibinfo {author}
  {\bibfnamefont {S.~L.}\ \bibnamefont {Shang}}, \bibinfo {author}
  {\bibfnamefont {D.~J.}\ \bibnamefont {Baek}}, \bibinfo {author}
  {\bibfnamefont {B.~H.}\ \bibnamefont {Goodge}}, \bibinfo {author}
  {\bibfnamefont {L.~F.}\ \bibnamefont {Kourkoutis}}, \bibinfo {author}
  {\bibfnamefont {Z.~K.}\ \bibnamefont {Liu}}, \bibinfo {author} {\bibfnamefont
  {K.~M.}\ \bibnamefont {Shen}},\ and\ \bibinfo {author} {\bibfnamefont
  {D.~G.}\ \bibnamefont {Schlom}},\ }\href {https://doi.org/10.1063/1.5023477}
  {\bibfield  {journal} {\bibinfo  {journal} {APL Mater.}\ }\textbf {\bibinfo
  {volume} {6}},\ \bibinfo {pages} {046101} (\bibinfo {year}
  {2018})}\BibitemShut {NoStop}%
\bibitem [{\citenamefont {Chambers}\ \emph {et~al.}(2002)\citenamefont
  {Chambers}, \citenamefont {Farrow}, \citenamefont {Maat}, \citenamefont
  {Toney}, \citenamefont {Folks}, \citenamefont {Catalano}, \citenamefont
  {Trainor},\ and\ \citenamefont {Brown}}]{Chambers2002}%
  \BibitemOpen
  \bibfield  {author} {\bibinfo {author} {\bibfnamefont {S.~A.}\ \bibnamefont
  {Chambers}}, \bibinfo {author} {\bibfnamefont {R.~F.}\ \bibnamefont
  {Farrow}}, \bibinfo {author} {\bibfnamefont {S.}~\bibnamefont {Maat}},
  \bibinfo {author} {\bibfnamefont {M.~F.}\ \bibnamefont {Toney}}, \bibinfo
  {author} {\bibfnamefont {L.}~\bibnamefont {Folks}}, \bibinfo {author}
  {\bibfnamefont {J.~G.}\ \bibnamefont {Catalano}}, \bibinfo {author}
  {\bibfnamefont {T.~P.}\ \bibnamefont {Trainor}},\ and\ \bibinfo {author}
  {\bibfnamefont {G.~E.}\ \bibnamefont {Brown}},\ }\href
  {https://doi.org/10.1016/S0304-8853(02)00039-2} {\bibfield  {journal}
  {\bibinfo  {journal} {J. Magn. Magn. Mater.}\ }\textbf {\bibinfo {volume}
  {246}},\ \bibinfo {pages} {124--139} (\bibinfo {year} {2002})}\BibitemShut
  {NoStop}%
\bibitem [{\citenamefont {Boris}\ \emph {et~al.}(2011)\citenamefont {Boris},
  \citenamefont {Matiks}, \citenamefont {Benckiser}, \citenamefont {Frano},
  \citenamefont {Popovich}, \citenamefont {Hinkov}, \citenamefont {Wochner},
  \citenamefont {Castro-Colin}, \citenamefont {Detemple}, \citenamefont
  {Malik}, \citenamefont {Bernhard}, \citenamefont {Prokscha}, \citenamefont
  {Suter}, \citenamefont {Salman}, \citenamefont {Morenzoni}, \citenamefont
  {Cristiani}, \citenamefont {Habermeier},\ and\ \citenamefont
  {Keimer}}]{Boris2011}%
  \BibitemOpen
  \bibfield  {author} {\bibinfo {author} {\bibfnamefont {A.~V.}\ \bibnamefont
  {Boris}}, \bibinfo {author} {\bibfnamefont {Y.}~\bibnamefont {Matiks}},
  \bibinfo {author} {\bibfnamefont {E.}~\bibnamefont {Benckiser}}, \bibinfo
  {author} {\bibfnamefont {A.}~\bibnamefont {Frano}}, \bibinfo {author}
  {\bibfnamefont {P.}~\bibnamefont {Popovich}}, \bibinfo {author}
  {\bibfnamefont {V.}~\bibnamefont {Hinkov}}, \bibinfo {author} {\bibfnamefont
  {P.}~\bibnamefont {Wochner}}, \bibinfo {author} {\bibfnamefont
  {M.}~\bibnamefont {Castro-Colin}}, \bibinfo {author} {\bibfnamefont
  {E.}~\bibnamefont {Detemple}}, \bibinfo {author} {\bibfnamefont {V.~K.}\
  \bibnamefont {Malik}}, \bibinfo {author} {\bibfnamefont {C.}~\bibnamefont
  {Bernhard}}, \bibinfo {author} {\bibfnamefont {T.}~\bibnamefont {Prokscha}},
  \bibinfo {author} {\bibfnamefont {A.}~\bibnamefont {Suter}}, \bibinfo
  {author} {\bibfnamefont {Z.}~\bibnamefont {Salman}}, \bibinfo {author}
  {\bibfnamefont {E.}~\bibnamefont {Morenzoni}}, \bibinfo {author}
  {\bibfnamefont {G.}~\bibnamefont {Cristiani}}, \bibinfo {author}
  {\bibfnamefont {H.~U.}\ \bibnamefont {Habermeier}},\ and\ \bibinfo {author}
  {\bibfnamefont {B.}~\bibnamefont {Keimer}},\ }\href
  {https://doi.org/10.1126/science.1202647} {\bibfield  {journal} {\bibinfo
  {journal} {Science}\ }\textbf {\bibinfo {volume} {332}},\ \bibinfo {pages}
  {937--940} (\bibinfo {year} {2011})}\BibitemShut {NoStop}%
\bibitem [{\citenamefont {Chakhalian}\ \emph {et~al.}(2006)\citenamefont
  {Chakhalian}, \citenamefont {Freeland}, \citenamefont {Srajer}, \citenamefont
  {Strempfer}, \citenamefont {Khaliullin}, \citenamefont {Cezar}, \citenamefont
  {Charlton}, \citenamefont {Dalgliesh}, \citenamefont {Bernhard},
  \citenamefont {Cristiani}, \citenamefont {Habermeier},\ and\ \citenamefont
  {Keimer}}]{Chakhalian2006}%
  \BibitemOpen
  \bibfield  {author} {\bibinfo {author} {\bibfnamefont {J.}~\bibnamefont
  {Chakhalian}}, \bibinfo {author} {\bibfnamefont {J.~W.}\ \bibnamefont
  {Freeland}}, \bibinfo {author} {\bibfnamefont {G.}~\bibnamefont {Srajer}},
  \bibinfo {author} {\bibfnamefont {J.}~\bibnamefont {Strempfer}}, \bibinfo
  {author} {\bibfnamefont {G.}~\bibnamefont {Khaliullin}}, \bibinfo {author}
  {\bibfnamefont {J.~C.}\ \bibnamefont {Cezar}}, \bibinfo {author}
  {\bibfnamefont {T.}~\bibnamefont {Charlton}}, \bibinfo {author}
  {\bibfnamefont {R.}~\bibnamefont {Dalgliesh}}, \bibinfo {author}
  {\bibfnamefont {C.}~\bibnamefont {Bernhard}}, \bibinfo {author}
  {\bibfnamefont {G.}~\bibnamefont {Cristiani}}, \bibinfo {author}
  {\bibfnamefont {H.~U.}\ \bibnamefont {Habermeier}},\ and\ \bibinfo {author}
  {\bibfnamefont {B.}~\bibnamefont {Keimer}},\ }\href
  {https://doi.org/10.1038/nphys272} {\bibfield  {journal} {\bibinfo  {journal}
  {Nat. Phys.}\ }\textbf {\bibinfo {volume} {2}},\ \bibinfo {pages} {244--248}
  (\bibinfo {year} {2006})}\BibitemShut {NoStop}%
\bibitem [{\citenamefont {Frano}\ \emph {et~al.}(2013)\citenamefont {Frano},
  \citenamefont {Schierle}, \citenamefont {Haverkort}, \citenamefont {Lu},
  \citenamefont {Wu}, \citenamefont {Blanco-Canosa}, \citenamefont {Nwankwo},
  \citenamefont {Boris}, \citenamefont {Wochner}, \citenamefont {Cristiani},
  \citenamefont {Habermeier}, \citenamefont {Logvenov}, \citenamefont {Hinkov},
  \citenamefont {Benckiser}, \citenamefont {Weschke},\ and\ \citenamefont
  {Keimer}}]{Frano2013}%
  \BibitemOpen
  \bibfield  {author} {\bibinfo {author} {\bibfnamefont {A.}~\bibnamefont
  {Frano}}, \bibinfo {author} {\bibfnamefont {E.}~\bibnamefont {Schierle}},
  \bibinfo {author} {\bibfnamefont {M.~W.}\ \bibnamefont {Haverkort}}, \bibinfo
  {author} {\bibfnamefont {Y.}~\bibnamefont {Lu}}, \bibinfo {author}
  {\bibfnamefont {M.}~\bibnamefont {Wu}}, \bibinfo {author} {\bibfnamefont
  {S.}~\bibnamefont {Blanco-Canosa}}, \bibinfo {author} {\bibfnamefont
  {U.}~\bibnamefont {Nwankwo}}, \bibinfo {author} {\bibfnamefont {A.~V.}\
  \bibnamefont {Boris}}, \bibinfo {author} {\bibfnamefont {P.}~\bibnamefont
  {Wochner}}, \bibinfo {author} {\bibfnamefont {G.}~\bibnamefont {Cristiani}},
  \bibinfo {author} {\bibfnamefont {H.~U.}\ \bibnamefont {Habermeier}},
  \bibinfo {author} {\bibfnamefont {G.}~\bibnamefont {Logvenov}}, \bibinfo
  {author} {\bibfnamefont {V.}~\bibnamefont {Hinkov}}, \bibinfo {author}
  {\bibfnamefont {E.}~\bibnamefont {Benckiser}}, \bibinfo {author}
  {\bibfnamefont {E.}~\bibnamefont {Weschke}},\ and\ \bibinfo {author}
  {\bibfnamefont {B.}~\bibnamefont {Keimer}},\ }\href
  {https://doi.org/10.1103/PhysRevLett.111.106804} {\bibfield  {journal}
  {\bibinfo  {journal} {Phys. Rev. Lett.}\ }\textbf {\bibinfo {volume} {111}},\
  \bibinfo {pages} {106804} (\bibinfo {year} {2013})}\BibitemShut {NoStop}%
\bibitem [{\citenamefont {Gao}\ \emph {et~al.}(1991)\citenamefont {Gao},
  \citenamefont {Aarnink}, \citenamefont {Gerritsma}, \citenamefont {Rijnders},
  \citenamefont {Rogalla}, \citenamefont {Hakkens}, \citenamefont {Coene},\
  and\ \citenamefont {Gijs}}]{Gao1991}%
  \BibitemOpen
  \bibfield  {author} {\bibinfo {author} {\bibfnamefont {J.}~\bibnamefont
  {Gao}}, \bibinfo {author} {\bibfnamefont {W.~A.~M.}\ \bibnamefont {Aarnink}},
  \bibinfo {author} {\bibfnamefont {G.~J.}\ \bibnamefont {Gerritsma}}, \bibinfo
  {author} {\bibfnamefont {A.~J. H.~M.}\ \bibnamefont {Rijnders}}, \bibinfo
  {author} {\bibfnamefont {H.}~\bibnamefont {Rogalla}}, \bibinfo {author}
  {\bibfnamefont {F.}~\bibnamefont {Hakkens}}, \bibinfo {author} {\bibfnamefont
  {W.}~\bibnamefont {Coene}},\ and\ \bibinfo {author} {\bibfnamefont
  {M.~A.~M.}\ \bibnamefont {Gijs}},\ }\href
  {https://doi.org/https://doi.org/10.1016/0921-4534(91)90495-K} {\bibfield
  {journal} {\bibinfo  {journal} {Physica C}\ }\textbf {\bibinfo {volume}
  {177}},\ \bibinfo {pages} {384--392} (\bibinfo {year} {1991})}\BibitemShut
  {NoStop}%
\bibitem [{\citenamefont {Henini}(2012)}]{Henini2012}%
  \BibitemOpen
  \bibfield  {author} {\bibinfo {author} {\bibfnamefont {M.}~\bibnamefont
  {Henini}},\ }\href@noop {} {\emph {\bibinfo {title} {Molecular Beam Epitaxy:
  From Research to Mass Production}}}\ (\bibinfo  {publisher} {Elsevier
  Science},\ \bibinfo {year} {2012})\BibitemShut {NoStop}%
\bibitem [{\citenamefont {Pfeiffer}\ and\ \citenamefont
  {West}(2003)}]{Pfeiffer2003}%
  \BibitemOpen
  \bibfield  {author} {\bibinfo {author} {\bibfnamefont {L.}~\bibnamefont
  {Pfeiffer}}\ and\ \bibinfo {author} {\bibfnamefont {K.~W.}\ \bibnamefont
  {West}},\ }\href {https://doi.org/10.1016/j.physe.2003.09.035} {\bibfield
  {journal} {\bibinfo  {journal} {Physica E}\ }\textbf {\bibinfo {volume}
  {20}},\ \bibinfo {pages} {57--64} (\bibinfo {year} {2003})}\BibitemShut
  {NoStop}%
\bibitem [{\citenamefont {Falson}\ \emph {et~al.}(2022)\citenamefont {Falson},
  \citenamefont {Sodemann}, \citenamefont {Skinner}, \citenamefont {Tabrea},
  \citenamefont {Kozuka}, \citenamefont {Tsukazaki}, \citenamefont {Kawasaki},
  \citenamefont {von Klitzing},\ and\ \citenamefont {Smet}}]{Falson2022}%
  \BibitemOpen
  \bibfield  {author} {\bibinfo {author} {\bibfnamefont {J.}~\bibnamefont
  {Falson}}, \bibinfo {author} {\bibfnamefont {I.}~\bibnamefont {Sodemann}},
  \bibinfo {author} {\bibfnamefont {B.}~\bibnamefont {Skinner}}, \bibinfo
  {author} {\bibfnamefont {D.}~\bibnamefont {Tabrea}}, \bibinfo {author}
  {\bibfnamefont {Y.}~\bibnamefont {Kozuka}}, \bibinfo {author} {\bibfnamefont
  {A.}~\bibnamefont {Tsukazaki}}, \bibinfo {author} {\bibfnamefont
  {M.}~\bibnamefont {Kawasaki}}, \bibinfo {author} {\bibfnamefont
  {K.}~\bibnamefont {von Klitzing}},\ and\ \bibinfo {author} {\bibfnamefont
  {J.~H.}\ \bibnamefont {Smet}},\ }\href
  {https://doi.org/10.1038/s41563-021-01166-1} {\bibfield  {journal} {\bibinfo
  {journal} {Nat. Mater.}\ }\textbf {\bibinfo {volume} {21}},\ \bibinfo {pages}
  {311--316} (\bibinfo {year} {2022})}\BibitemShut {NoStop}%
\bibitem [{\citenamefont {Hellman}\ and\ \citenamefont
  {Hartford}(1990)}]{Hellman1990}%
  \BibitemOpen
  \bibfield  {author} {\bibinfo {author} {\bibfnamefont {E.~S.}\ \bibnamefont
  {Hellman}}\ and\ \bibinfo {author} {\bibfnamefont {E.~H.}\ \bibnamefont
  {Hartford}},\ }\href {https://doi.org/10.1116/1.585064} {\bibfield  {journal}
  {\bibinfo  {journal} {J. Vac. Sci. Technol. B}\ }\textbf {\bibinfo {volume}
  {8}},\ \bibinfo {pages} {332--335} (\bibinfo {year} {1990})}\BibitemShut
  {NoStop}%
\bibitem [{\citenamefont {Theis}\ \emph
  {et~al.}(1998{\natexlab{a}})\citenamefont {Theis}, \citenamefont {Yeh},
  \citenamefont {Schlom}, \citenamefont {Hawley},\ and\ \citenamefont
  {Brown}}]{Theis1998}%
  \BibitemOpen
  \bibfield  {author} {\bibinfo {author} {\bibfnamefont {C.~D.}\ \bibnamefont
  {Theis}}, \bibinfo {author} {\bibfnamefont {J.}~\bibnamefont {Yeh}}, \bibinfo
  {author} {\bibfnamefont {D.~G.}\ \bibnamefont {Schlom}}, \bibinfo {author}
  {\bibfnamefont {M.~E.}\ \bibnamefont {Hawley}},\ and\ \bibinfo {author}
  {\bibfnamefont {G.~W.}\ \bibnamefont {Brown}},\ }\href
  {https://doi.org/10.1016/S0040-6090(98)00507-0} {\bibfield  {journal}
  {\bibinfo  {journal} {Thin Solid Films}\ }\textbf {\bibinfo {volume} {325}},\
  \bibinfo {pages} {107--114} (\bibinfo {year}
  {1998}{\natexlab{a}})}\BibitemShut {NoStop}%
\bibitem [{\citenamefont {Theis}\ \emph
  {et~al.}(1998{\natexlab{b}})\citenamefont {Theis}, \citenamefont {Yeh},
  \citenamefont {Schlom}, \citenamefont {Hawley}, \citenamefont {Brown},
  \citenamefont {Jiang},\ and\ \citenamefont {Pan}}]{Theis1998b}%
  \BibitemOpen
  \bibfield  {author} {\bibinfo {author} {\bibfnamefont {C.}~\bibnamefont
  {Theis}}, \bibinfo {author} {\bibfnamefont {J.}~\bibnamefont {Yeh}}, \bibinfo
  {author} {\bibfnamefont {D.}~\bibnamefont {Schlom}}, \bibinfo {author}
  {\bibfnamefont {M.}~\bibnamefont {Hawley}}, \bibinfo {author} {\bibfnamefont
  {G.}~\bibnamefont {Brown}}, \bibinfo {author} {\bibfnamefont
  {J.}~\bibnamefont {Jiang}},\ and\ \bibinfo {author} {\bibfnamefont
  {X.}~\bibnamefont {Pan}},\ }\href {https://doi.org/10.1063/1.121468}
  {\bibfield  {journal} {\bibinfo  {journal} {Appl. Phys. Lett.}\ }\textbf
  {\bibinfo {volume} {72}},\ \bibinfo {pages} {2817} (\bibinfo {year}
  {1998}{\natexlab{b}})}\BibitemShut {NoStop}%
\bibitem [{\citenamefont {Jalan}\ \emph {et~al.}(2009)\citenamefont {Jalan},
  \citenamefont {Engel-Herbert}, \citenamefont {Wright},\ and\ \citenamefont
  {Stemmer}}]{Jalan2009}%
  \BibitemOpen
  \bibfield  {author} {\bibinfo {author} {\bibfnamefont {B.}~\bibnamefont
  {Jalan}}, \bibinfo {author} {\bibfnamefont {R.}~\bibnamefont
  {Engel-Herbert}}, \bibinfo {author} {\bibfnamefont {N.~J.}\ \bibnamefont
  {Wright}},\ and\ \bibinfo {author} {\bibfnamefont {S.}~\bibnamefont
  {Stemmer}},\ }\href {https://doi.org/10.1116/1.3106610} {\bibfield  {journal}
  {\bibinfo  {journal} {J. Vac. Sci. Technol. A}\ }\textbf {\bibinfo {volume}
  {27}},\ \bibinfo {pages} {461--464} (\bibinfo {year} {2009})}\BibitemShut
  {NoStop}%
\bibitem [{\citenamefont {Brahlek}\ \emph {et~al.}(2018)\citenamefont
  {Brahlek}, \citenamefont {Gupta}, \citenamefont {Lapano}, \citenamefont
  {Roth}, \citenamefont {Zhang}, \citenamefont {Zhang}, \citenamefont
  {Haislmaier},\ and\ \citenamefont {Engel-Herbert}}]{Brahlek2018}%
  \BibitemOpen
  \bibfield  {author} {\bibinfo {author} {\bibfnamefont {M.}~\bibnamefont
  {Brahlek}}, \bibinfo {author} {\bibfnamefont {A.~S.}\ \bibnamefont {Gupta}},
  \bibinfo {author} {\bibfnamefont {J.}~\bibnamefont {Lapano}}, \bibinfo
  {author} {\bibfnamefont {J.}~\bibnamefont {Roth}}, \bibinfo {author}
  {\bibfnamefont {H.~T.}\ \bibnamefont {Zhang}}, \bibinfo {author}
  {\bibfnamefont {L.}~\bibnamefont {Zhang}}, \bibinfo {author} {\bibfnamefont
  {R.}~\bibnamefont {Haislmaier}},\ and\ \bibinfo {author} {\bibfnamefont
  {R.}~\bibnamefont {Engel-Herbert}},\ }\href
  {https://doi.org/10.1002/adfm.201702772} {\bibfield  {journal} {\bibinfo
  {journal} {Adv. Func. Mater.}\ }\textbf {\bibinfo {volume} {28}},\ \bibinfo
  {pages} {1702772} (\bibinfo {year} {2018})}\BibitemShut {NoStop}%
\bibitem [{\citenamefont {Marquardt}, \citenamefont {Ashmore},\ and\
  \citenamefont {Cann}(2006)}]{Marquardt2006}%
  \BibitemOpen
  \bibfield  {author} {\bibinfo {author} {\bibfnamefont {M.~A.}\ \bibnamefont
  {Marquardt}}, \bibinfo {author} {\bibfnamefont {N.~A.}\ \bibnamefont
  {Ashmore}},\ and\ \bibinfo {author} {\bibfnamefont {D.~P.}\ \bibnamefont
  {Cann}},\ }\href {https://doi.org/10.1016/j.tsf.2005.08.316} {\bibfield
  {journal} {\bibinfo  {journal} {Thin Solid Films}\ }\textbf {\bibinfo
  {volume} {496}},\ \bibinfo {pages} {146--156} (\bibinfo {year}
  {2006})}\BibitemShut {NoStop}%
\bibitem [{\citenamefont {Mackenzie}(2017)}]{Mackenzie2017}%
  \BibitemOpen
  \bibfield  {author} {\bibinfo {author} {\bibfnamefont {A.~P.}\ \bibnamefont
  {Mackenzie}},\ }\href {https://doi.org/10.1088/1361-6633/aa50e5} {\bibfield
  {journal} {\bibinfo  {journal} {Rep. Prog. Phys.}\ }\textbf {\bibinfo
  {volume} {80}},\ \bibinfo {pages} {032501} (\bibinfo {year}
  {2017})}\BibitemShut {NoStop}%
\bibitem [{\citenamefont {Hicks}\ \emph {et~al.}(2012)\citenamefont {Hicks},
  \citenamefont {Gibbs}, \citenamefont {Mackenzie}, \citenamefont {Takatsu},
  \citenamefont {Maeno},\ and\ \citenamefont {Yelland}}]{Hicks2012}%
  \BibitemOpen
  \bibfield  {author} {\bibinfo {author} {\bibfnamefont {C.~W.}\ \bibnamefont
  {Hicks}}, \bibinfo {author} {\bibfnamefont {A.~S.}\ \bibnamefont {Gibbs}},
  \bibinfo {author} {\bibfnamefont {A.~P.}\ \bibnamefont {Mackenzie}}, \bibinfo
  {author} {\bibfnamefont {H.}~\bibnamefont {Takatsu}}, \bibinfo {author}
  {\bibfnamefont {Y.}~\bibnamefont {Maeno}},\ and\ \bibinfo {author}
  {\bibfnamefont {E.~A.}\ \bibnamefont {Yelland}},\ }\href
  {https://doi.org/10.1103/PhysRevLett.109.116401} {\bibfield  {journal}
  {\bibinfo  {journal} {Phys. Rev. Lett.}\ }\textbf {\bibinfo {volume} {109}},\
  \bibinfo {pages} {116401} (\bibinfo {year} {2012})}\BibitemShut {NoStop}%
\bibitem [{\citenamefont {Moll}\ \emph {et~al.}(2016)\citenamefont {Moll},
  \citenamefont {Kushwaha}, \citenamefont {Nandi}, \citenamefont {Schmidt},\
  and\ \citenamefont {Mackenzie}}]{Moll2016}%
  \BibitemOpen
  \bibfield  {author} {\bibinfo {author} {\bibfnamefont {P.~J.}\ \bibnamefont
  {Moll}}, \bibinfo {author} {\bibfnamefont {P.}~\bibnamefont {Kushwaha}},
  \bibinfo {author} {\bibfnamefont {N.}~\bibnamefont {Nandi}}, \bibinfo
  {author} {\bibfnamefont {B.}~\bibnamefont {Schmidt}},\ and\ \bibinfo {author}
  {\bibfnamefont {A.~P.}\ \bibnamefont {Mackenzie}},\ }\href
  {https://doi.org/10.1126/science.aac8385} {\bibfield  {journal} {\bibinfo
  {journal} {Science}\ }\textbf {\bibinfo {volume} {351}},\ \bibinfo {pages}
  {1061} (\bibinfo {year} {2016})}\BibitemShut {NoStop}%
\bibitem [{\citenamefont {Mazzola}\ \emph {et~al.}(2017)\citenamefont
  {Mazzola}, \citenamefont {Sunko}, \citenamefont {Khim}, \citenamefont
  {Rosner}, \citenamefont {Kushwaha}, \citenamefont {Clark}, \citenamefont
  {Bawden}, \citenamefont {Markovi{\'{c}}}, \citenamefont {Kim}, \citenamefont
  {Hoesch}, \citenamefont {Mackenzie},\ and\ \citenamefont
  {King}}]{Mazzola2017}%
  \BibitemOpen
  \bibfield  {author} {\bibinfo {author} {\bibfnamefont {F.}~\bibnamefont
  {Mazzola}}, \bibinfo {author} {\bibfnamefont {V.}~\bibnamefont {Sunko}},
  \bibinfo {author} {\bibfnamefont {S.}~\bibnamefont {Khim}}, \bibinfo {author}
  {\bibfnamefont {H.}~\bibnamefont {Rosner}}, \bibinfo {author} {\bibfnamefont
  {P.}~\bibnamefont {Kushwaha}}, \bibinfo {author} {\bibfnamefont {O.~J.}\
  \bibnamefont {Clark}}, \bibinfo {author} {\bibfnamefont {L.}~\bibnamefont
  {Bawden}}, \bibinfo {author} {\bibfnamefont {I.}~\bibnamefont
  {Markovi{\'{c}}}}, \bibinfo {author} {\bibfnamefont {T.~K.}\ \bibnamefont
  {Kim}}, \bibinfo {author} {\bibfnamefont {M.}~\bibnamefont {Hoesch}},
  \bibinfo {author} {\bibfnamefont {A.~P.}\ \bibnamefont {Mackenzie}},\ and\
  \bibinfo {author} {\bibfnamefont {P.~D.~C.}\ \bibnamefont {King}},\ }\href
  {https://doi.org/10.1073/pnas.1811873115} {\bibfield  {journal} {\bibinfo
  {journal} {PNAS}\ }\textbf {\bibinfo {volume} {115}},\ \bibinfo {pages}
  {12956} (\bibinfo {year} {2017})}\BibitemShut {NoStop}%
\bibitem [{\citenamefont {Coldea}\ \emph {et~al.}(2014)\citenamefont {Coldea},
  \citenamefont {Seabra}, \citenamefont {McCollam}, \citenamefont {Carrington},
  \citenamefont {Malone}, \citenamefont {Bangura}, \citenamefont {Vignolles},
  \citenamefont {{Van Rhee}}, \citenamefont {McDonald}, \citenamefont
  {S{\"{o}}rgel}, \citenamefont {Jansen}, \citenamefont {Shannon},\ and\
  \citenamefont {Coldea}}]{Coldea2014}%
  \BibitemOpen
  \bibfield  {author} {\bibinfo {author} {\bibfnamefont {A.~I.}\ \bibnamefont
  {Coldea}}, \bibinfo {author} {\bibfnamefont {L.}~\bibnamefont {Seabra}},
  \bibinfo {author} {\bibfnamefont {A.}~\bibnamefont {McCollam}}, \bibinfo
  {author} {\bibfnamefont {A.}~\bibnamefont {Carrington}}, \bibinfo {author}
  {\bibfnamefont {L.}~\bibnamefont {Malone}}, \bibinfo {author} {\bibfnamefont
  {A.~F.}\ \bibnamefont {Bangura}}, \bibinfo {author} {\bibfnamefont
  {D.}~\bibnamefont {Vignolles}}, \bibinfo {author} {\bibfnamefont {P.~G.}\
  \bibnamefont {{Van Rhee}}}, \bibinfo {author} {\bibfnamefont {R.~D.}\
  \bibnamefont {McDonald}}, \bibinfo {author} {\bibfnamefont {T.}~\bibnamefont
  {S{\"{o}}rgel}}, \bibinfo {author} {\bibfnamefont {M.}~\bibnamefont
  {Jansen}}, \bibinfo {author} {\bibfnamefont {N.}~\bibnamefont {Shannon}},\
  and\ \bibinfo {author} {\bibfnamefont {R.}~\bibnamefont {Coldea}},\ }\href
  {https://doi.org/10.1103/PhysRevB.90.020401} {\bibfield  {journal} {\bibinfo
  {journal} {Phys. Rev. B}\ }\textbf {\bibinfo {volume} {90}},\ \bibinfo
  {pages} {020401} (\bibinfo {year} {2014})}\BibitemShut {NoStop}%
\bibitem [{\citenamefont {Sunko}\ \emph {et~al.}(2020)\citenamefont {Sunko},
  \citenamefont {Mcguinness}, \citenamefont {Chang}, \citenamefont {Zhakina},
  \citenamefont {Khim}, \citenamefont {Dreyer}, \citenamefont {Konczykowski},
  \citenamefont {Borrmann}, \citenamefont {Moll}, \citenamefont {K{\"{o}}nig},
  \citenamefont {Muller},\ and\ \citenamefont {Mackenzie}}]{Sunko2020}%
  \BibitemOpen
  \bibfield  {author} {\bibinfo {author} {\bibfnamefont {V.}~\bibnamefont
  {Sunko}}, \bibinfo {author} {\bibfnamefont {P.~H.}\ \bibnamefont
  {Mcguinness}}, \bibinfo {author} {\bibfnamefont {C.~S.}\ \bibnamefont
  {Chang}}, \bibinfo {author} {\bibfnamefont {E.}~\bibnamefont {Zhakina}},
  \bibinfo {author} {\bibfnamefont {S.}~\bibnamefont {Khim}}, \bibinfo {author}
  {\bibfnamefont {C.~E.}\ \bibnamefont {Dreyer}}, \bibinfo {author}
  {\bibfnamefont {M.}~\bibnamefont {Konczykowski}}, \bibinfo {author}
  {\bibfnamefont {H.}~\bibnamefont {Borrmann}}, \bibinfo {author}
  {\bibfnamefont {P.~J.}\ \bibnamefont {Moll}}, \bibinfo {author}
  {\bibfnamefont {M.}~\bibnamefont {K{\"{o}}nig}}, \bibinfo {author}
  {\bibfnamefont {D.~A.}\ \bibnamefont {Muller}},\ and\ \bibinfo {author}
  {\bibfnamefont {A.~P.}\ \bibnamefont {Mackenzie}},\ }\href
  {https://doi.org/10.1103/PhysRevX.10.021018} {\bibfield  {journal} {\bibinfo
  {journal} {Phys. Rev. X}\ }\textbf {\bibinfo {volume} {10}},\ \bibinfo
  {pages} {021018} (\bibinfo {year} {2020})}\BibitemShut {NoStop}%
\bibitem [{\citenamefont {Scanlon}\ and\ \citenamefont
  {Watson}(2011)}]{Scanlon2011}%
  \BibitemOpen
  \bibfield  {author} {\bibinfo {author} {\bibfnamefont {D.~O.}\ \bibnamefont
  {Scanlon}}\ and\ \bibinfo {author} {\bibfnamefont {G.~W.}\ \bibnamefont
  {Watson}},\ }\href {https://doi.org/10.1039/c0jm03852k} {\bibfield  {journal}
  {\bibinfo  {journal} {J. Mat. Chem.}\ }\textbf {\bibinfo {volume} {21}},\
  \bibinfo {pages} {3655--3663} (\bibinfo {year} {2011})}\BibitemShut {NoStop}%
\bibitem [{\citenamefont {Chiba}, \citenamefont {Kawashima},\ and\
  \citenamefont {Washio}(2017)}]{Chiba2017}%
  \BibitemOpen
  \bibfield  {author} {\bibinfo {author} {\bibfnamefont {H.}~\bibnamefont
  {Chiba}}, \bibinfo {author} {\bibfnamefont {T.}~\bibnamefont {Kawashima}},\
  and\ \bibinfo {author} {\bibfnamefont {K.}~\bibnamefont {Washio}},\ }\href
  {https://doi.org/10.1016/j.mssp.2016.10.013} {\bibfield  {journal} {\bibinfo
  {journal} {Mat. Sci. Semicond. Proc.}\ }\textbf {\bibinfo {volume} {70}},\
  \bibinfo {pages} {234--238} (\bibinfo {year} {2017})}\BibitemShut {NoStop}%
\bibitem [{\citenamefont {Chiba}\ \emph {et~al.}(2018)\citenamefont {Chiba},
  \citenamefont {Hosaka}, \citenamefont {Kawashima},\ and\ \citenamefont
  {Washio}}]{Chiba2018}%
  \BibitemOpen
  \bibfield  {author} {\bibinfo {author} {\bibfnamefont {H.}~\bibnamefont
  {Chiba}}, \bibinfo {author} {\bibfnamefont {N.}~\bibnamefont {Hosaka}},
  \bibinfo {author} {\bibfnamefont {T.}~\bibnamefont {Kawashima}},\ and\
  \bibinfo {author} {\bibfnamefont {K.}~\bibnamefont {Washio}},\ }\href
  {https://doi.org/10.1016/j.tsf.2017.11.028} {\bibfield  {journal} {\bibinfo
  {journal} {Thin Solid Films}\ }\textbf {\bibinfo {volume} {652}},\ \bibinfo
  {pages} {16--22} (\bibinfo {year} {2018})}\BibitemShut {NoStop}%
\bibitem [{\citenamefont {Tripathi}\ and\ \citenamefont
  {Karppinen}(2017)}]{Tripathi2017}%
  \BibitemOpen
  \bibfield  {author} {\bibinfo {author} {\bibfnamefont {T.~S.}\ \bibnamefont
  {Tripathi}}\ and\ \bibinfo {author} {\bibfnamefont {M.}~\bibnamefont
  {Karppinen}},\ }\href {https://doi.org/10.1002/aelm.201600341} {\bibfield
  {journal} {\bibinfo  {journal} {Adv. Electron. Mater.}\ }\textbf {\bibinfo
  {volume} {3}},\ \bibinfo {pages} {1600341} (\bibinfo {year}
  {2017})}\BibitemShut {NoStop}%
\bibitem [{\citenamefont {Kimura}, \citenamefont {Lashley},\ and\ \citenamefont
  {Ramirez}(2006)}]{Kimura2006}%
  \BibitemOpen
  \bibfield  {author} {\bibinfo {author} {\bibfnamefont {T.}~\bibnamefont
  {Kimura}}, \bibinfo {author} {\bibfnamefont {J.~C.}\ \bibnamefont
  {Lashley}},\ and\ \bibinfo {author} {\bibfnamefont {A.~P.}\ \bibnamefont
  {Ramirez}},\ }\href {https://doi.org/10.1103/PhysRevB.73.220401} {\bibfield
  {journal} {\bibinfo  {journal} {Phys. Rev. B}\ }\textbf {\bibinfo {volume}
  {73}},\ \bibinfo {pages} {220401} (\bibinfo {year} {2006})}\BibitemShut
  {NoStop}%
\bibitem [{\citenamefont {Seki}, \citenamefont {Onose},\ and\ \citenamefont
  {Tokura}(2008)}]{Seki2008}%
  \BibitemOpen
  \bibfield  {author} {\bibinfo {author} {\bibfnamefont {S.}~\bibnamefont
  {Seki}}, \bibinfo {author} {\bibfnamefont {Y.}~\bibnamefont {Onose}},\ and\
  \bibinfo {author} {\bibfnamefont {Y.}~\bibnamefont {Tokura}},\ }\href
  {https://doi.org/10.1103/PhysRevLett.101.067204} {\bibfield  {journal}
  {\bibinfo  {journal} {Phys. Rev. Lett.}\ }\textbf {\bibinfo {volume} {101}},\
  \bibinfo {pages} {067204} (\bibinfo {year} {2008})}\BibitemShut {NoStop}%
\bibitem [{\citenamefont {Poienar}\ \emph {et~al.}(2010)\citenamefont
  {Poienar}, \citenamefont {Damay}, \citenamefont {Martin}, \citenamefont
  {Robert},\ and\ \citenamefont {Petit}}]{Poienar2010}%
  \BibitemOpen
  \bibfield  {author} {\bibinfo {author} {\bibfnamefont {M.}~\bibnamefont
  {Poienar}}, \bibinfo {author} {\bibfnamefont {F.}~\bibnamefont {Damay}},
  \bibinfo {author} {\bibfnamefont {C.}~\bibnamefont {Martin}}, \bibinfo
  {author} {\bibfnamefont {J.}~\bibnamefont {Robert}},\ and\ \bibinfo {author}
  {\bibfnamefont {S.}~\bibnamefont {Petit}},\ }\href
  {https://doi.org/10.1103/PhysRevB.81.104411} {\bibfield  {journal} {\bibinfo
  {journal} {Phys. Rev. B}\ }\textbf {\bibinfo {volume} {81}},\ \bibinfo
  {pages} {104411} (\bibinfo {year} {2010})}\BibitemShut {NoStop}%
\bibitem [{\citenamefont {Frontzek}\ \emph {et~al.}(2011)\citenamefont
  {Frontzek}, \citenamefont {Haraldsen}, \citenamefont {Podlesnyak},
  \citenamefont {Matsuda}, \citenamefont {Christianson}, \citenamefont
  {Fishman}, \citenamefont {Sefat}, \citenamefont {Qiu}, \citenamefont
  {Copley}, \citenamefont {Barilo}, \citenamefont {Shiryaev},\ and\
  \citenamefont {Ehlers}}]{Frontzek2011}%
  \BibitemOpen
  \bibfield  {author} {\bibinfo {author} {\bibfnamefont {M.}~\bibnamefont
  {Frontzek}}, \bibinfo {author} {\bibfnamefont {J.~T.}\ \bibnamefont
  {Haraldsen}}, \bibinfo {author} {\bibfnamefont {A.}~\bibnamefont
  {Podlesnyak}}, \bibinfo {author} {\bibfnamefont {M.}~\bibnamefont {Matsuda}},
  \bibinfo {author} {\bibfnamefont {A.~D.}\ \bibnamefont {Christianson}},
  \bibinfo {author} {\bibfnamefont {R.~S.}\ \bibnamefont {Fishman}}, \bibinfo
  {author} {\bibfnamefont {A.~S.}\ \bibnamefont {Sefat}}, \bibinfo {author}
  {\bibfnamefont {Y.}~\bibnamefont {Qiu}}, \bibinfo {author} {\bibfnamefont
  {J.~R.}\ \bibnamefont {Copley}}, \bibinfo {author} {\bibfnamefont
  {S.}~\bibnamefont {Barilo}}, \bibinfo {author} {\bibfnamefont {S.~V.}\
  \bibnamefont {Shiryaev}},\ and\ \bibinfo {author} {\bibfnamefont
  {G.}~\bibnamefont {Ehlers}},\ }\href
  {https://doi.org/10.1103/PhysRevB.84.094448} {\bibfield  {journal} {\bibinfo
  {journal} {Phys. Rev. B}\ }\textbf {\bibinfo {volume} {84}},\ \bibinfo
  {pages} {094448} (\bibinfo {year} {2011})}\BibitemShut {NoStop}%
\bibitem [{\citenamefont {Ok}\ \emph {et~al.}(2020{\natexlab{a}})\citenamefont
  {Ok}, \citenamefont {Brahlek}, \citenamefont {Choi}, \citenamefont
  {Roccapriore}, \citenamefont {Chisholm}, \citenamefont {Kim}, \citenamefont
  {Sohn}, \citenamefont {Skoropata}, \citenamefont {Yoon}, \citenamefont {Kim},
  \citenamefont {Lee},\ and\ \citenamefont {Nyung}}]{Ok2020}%
  \BibitemOpen
  \bibfield  {author} {\bibinfo {author} {\bibfnamefont {J.~M.}\ \bibnamefont
  {Ok}}, \bibinfo {author} {\bibfnamefont {M.}~\bibnamefont {Brahlek}},
  \bibinfo {author} {\bibfnamefont {W.~S.}\ \bibnamefont {Choi}}, \bibinfo
  {author} {\bibfnamefont {K.~M.}\ \bibnamefont {Roccapriore}}, \bibinfo
  {author} {\bibfnamefont {M.~F.}\ \bibnamefont {Chisholm}}, \bibinfo {author}
  {\bibfnamefont {S.}~\bibnamefont {Kim}}, \bibinfo {author} {\bibfnamefont
  {C.}~\bibnamefont {Sohn}}, \bibinfo {author} {\bibfnamefont {E.}~\bibnamefont
  {Skoropata}}, \bibinfo {author} {\bibfnamefont {S.}~\bibnamefont {Yoon}},
  \bibinfo {author} {\bibfnamefont {J.~S.}\ \bibnamefont {Kim}}, \bibinfo
  {author} {\bibfnamefont {H.~N.}\ \bibnamefont {Lee}},\ and\ \bibinfo {author}
  {\bibfnamefont {H.}~\bibnamefont {Nyung}},\ }\href
  {https://doi.org/10.1063/1.5144743} {\bibfield  {journal} {\bibinfo
  {journal} {APL Mater.}\ }\textbf {\bibinfo {volume} {8}},\ \bibinfo {pages}
  {051104} (\bibinfo {year} {2020}{\natexlab{a}})}\BibitemShut {NoStop}%
\bibitem [{\citenamefont {Shin}\ \emph {et~al.}(2012)\citenamefont {Shin},
  \citenamefont {Foord}, \citenamefont {Egdell},\ and\ \citenamefont
  {Walsh}}]{Shin2012}%
  \BibitemOpen
  \bibfield  {author} {\bibinfo {author} {\bibfnamefont {D.}~\bibnamefont
  {Shin}}, \bibinfo {author} {\bibfnamefont {J.~S.}\ \bibnamefont {Foord}},
  \bibinfo {author} {\bibfnamefont {R.~G.}\ \bibnamefont {Egdell}},\ and\
  \bibinfo {author} {\bibfnamefont {A.}~\bibnamefont {Walsh}},\ }\href
  {https://doi.org/10.1063/1.4768726} {\bibfield  {journal} {\bibinfo
  {journal} {J. Appl. Phys.}\ }\textbf {\bibinfo {volume} {112}},\ \bibinfo
  {pages} {113718} (\bibinfo {year} {2012})}\BibitemShut {NoStop}%
\bibitem [{\citenamefont {Brahlek}\ \emph {et~al.}(2019)\citenamefont
  {Brahlek}, \citenamefont {Rimal}, \citenamefont {Ok}, \citenamefont
  {Mukherjee}, \citenamefont {Mazza}, \citenamefont {Lu}, \citenamefont {Lee},
  \citenamefont {Ward}, \citenamefont {Unocic}, \citenamefont {Eres},\ and\
  \citenamefont {Oh}}]{Brahlek2019}%
  \BibitemOpen
  \bibfield  {author} {\bibinfo {author} {\bibfnamefont {M.}~\bibnamefont
  {Brahlek}}, \bibinfo {author} {\bibfnamefont {G.}~\bibnamefont {Rimal}},
  \bibinfo {author} {\bibfnamefont {J.~M.}\ \bibnamefont {Ok}}, \bibinfo
  {author} {\bibfnamefont {D.}~\bibnamefont {Mukherjee}}, \bibinfo {author}
  {\bibfnamefont {A.~R.}\ \bibnamefont {Mazza}}, \bibinfo {author}
  {\bibfnamefont {Q.}~\bibnamefont {Lu}}, \bibinfo {author} {\bibfnamefont
  {H.~N.}\ \bibnamefont {Lee}}, \bibinfo {author} {\bibfnamefont {T.~Z.}\
  \bibnamefont {Ward}}, \bibinfo {author} {\bibfnamefont {R.~R.}\ \bibnamefont
  {Unocic}}, \bibinfo {author} {\bibfnamefont {G.}~\bibnamefont {Eres}},\ and\
  \bibinfo {author} {\bibfnamefont {S.}~\bibnamefont {Oh}},\ }\href
  {https://doi.org/10.1103/PhysRevMat..3.093401} {\bibfield  {journal}
  {\bibinfo  {journal} {Phys. Rev. Mat.}\ }\textbf {\bibinfo {volume} {3}},\
  \bibinfo {pages} {093401} (\bibinfo {year} {2019})}\BibitemShut {NoStop}%
\bibitem [{\citenamefont {Sun}\ \emph {et~al.}(2019)\citenamefont {Sun},
  \citenamefont {Barone}, \citenamefont {Chang}, \citenamefont {Muller},
  \citenamefont {Holtz},\ and\ \citenamefont {Paik}}]{Sun2019}%
  \BibitemOpen
  \bibfield  {author} {\bibinfo {author} {\bibfnamefont {J.}~\bibnamefont
  {Sun}}, \bibinfo {author} {\bibfnamefont {M.~R.}\ \bibnamefont {Barone}},
  \bibinfo {author} {\bibfnamefont {C.~S.}\ \bibnamefont {Chang}}, \bibinfo
  {author} {\bibfnamefont {D.~A.}\ \bibnamefont {Muller}}, \bibinfo {author}
  {\bibfnamefont {M.~E.}\ \bibnamefont {Holtz}},\ and\ \bibinfo {author}
  {\bibfnamefont {H.}~\bibnamefont {Paik}},\ }\href
  {https://doi.org/10.1063/1.5130627} {\bibfield  {journal} {\bibinfo
  {journal} {APL Mater.}\ }\textbf {\bibinfo {volume} {7}},\ \bibinfo {pages}
  {121112} (\bibinfo {year} {2019})}\BibitemShut {NoStop}%
\bibitem [{\citenamefont {Bracht}(2004)}]{Bracht2004}%
  \BibitemOpen
  \bibfield  {author} {\bibinfo {author} {\bibfnamefont {H.}~\bibnamefont
  {Bracht}},\ }\href {https://doi.org/10.1016/j.mssp.2004.06.001} {\bibfield
  {journal} {\bibinfo  {journal} {Mater. Sci. Semicond. Process.}\ }\textbf
  {\bibinfo {volume} {7}},\ \bibinfo {pages} {113--124} (\bibinfo {year}
  {2004})}\BibitemShut {NoStop}%
\bibitem [{\citenamefont {Doremus}(2006)}]{Doremus2006}%
  \BibitemOpen
  \bibfield  {author} {\bibinfo {author} {\bibfnamefont {R.~H.}\ \bibnamefont
  {Doremus}},\ }\href {https://doi.org/10.1063/1.2393012} {\bibfield  {journal}
  {\bibinfo  {journal} {J. Appl. Phys.}\ }\textbf {\bibinfo {volume} {100}},\
  \bibinfo {pages} {101301} (\bibinfo {year} {2006})}\BibitemShut {NoStop}%
\bibitem [{\citenamefont {O'Sullivan}\ \emph {et~al.}(2010)\citenamefont
  {O'Sullivan}, \citenamefont {Stamenov}, \citenamefont {Alaria}, \citenamefont
  {Venkatesan},\ and\ \citenamefont {Coey}}]{OSullivan2010}%
  \BibitemOpen
  \bibfield  {author} {\bibinfo {author} {\bibfnamefont {M.}~\bibnamefont
  {O'Sullivan}}, \bibinfo {author} {\bibfnamefont {P.}~\bibnamefont
  {Stamenov}}, \bibinfo {author} {\bibfnamefont {J.}~\bibnamefont {Alaria}},
  \bibinfo {author} {\bibfnamefont {M.}~\bibnamefont {Venkatesan}},\ and\
  \bibinfo {author} {\bibfnamefont {J.~M.}\ \bibnamefont {Coey}},\ }\href
  {https://doi.org/10.1088/1742-6596/200/5/052021} {\bibfield  {journal}
  {\bibinfo  {journal} {J. Phys. Conf. Ser.}\ }\textbf {\bibinfo {volume}
  {200}},\ \bibinfo {pages} {6--10} (\bibinfo {year} {2010})}\BibitemShut
  {NoStop}%
\bibitem [{\citenamefont {Ok}\ \emph {et~al.}(2020{\natexlab{b}})\citenamefont
  {Ok}, \citenamefont {Yoon}, \citenamefont {Lupini}, \citenamefont {Ganesh},
  \citenamefont {Chisholm},\ and\ \citenamefont {Lee}}]{Ok2020b}%
  \BibitemOpen
  \bibfield  {author} {\bibinfo {author} {\bibfnamefont {J.~M.}\ \bibnamefont
  {Ok}}, \bibinfo {author} {\bibfnamefont {S.}~\bibnamefont {Yoon}}, \bibinfo
  {author} {\bibfnamefont {A.~R.}\ \bibnamefont {Lupini}}, \bibinfo {author}
  {\bibfnamefont {P.}~\bibnamefont {Ganesh}}, \bibinfo {author} {\bibfnamefont
  {M.~F.}\ \bibnamefont {Chisholm}},\ and\ \bibinfo {author} {\bibfnamefont
  {H.~N.}\ \bibnamefont {Lee}},\ }\href
  {https://doi.org/10.1038/s41598-020-68275-w} {\bibfield  {journal} {\bibinfo
  {journal} {Sci. Rep.}\ }\textbf {\bibinfo {volume} {10}},\ \bibinfo {pages}
  {11375} (\bibinfo {year} {2020}{\natexlab{b}})}\BibitemShut {NoStop}%
\bibitem [{\citenamefont {Poulston}\ \emph {et~al.}(1996)\citenamefont
  {Poulston}, \citenamefont {Parlett}, \citenamefont {Stone},\ and\
  \citenamefont {Bowker}}]{Poulston1996}%
  \BibitemOpen
  \bibfield  {author} {\bibinfo {author} {\bibfnamefont {S.}~\bibnamefont
  {Poulston}}, \bibinfo {author} {\bibfnamefont {P.~M.}\ \bibnamefont
  {Parlett}}, \bibinfo {author} {\bibfnamefont {P.}~\bibnamefont {Stone}},\
  and\ \bibinfo {author} {\bibfnamefont {M.}~\bibnamefont {Bowker}},\ }\href
  {https://doi.org/10.1002/(SICI)1096-9918(199611)24:12<811::AID-SIA191>3.0.CO;2-Z}
  {\bibfield  {journal} {\bibinfo  {journal} {Surf. Interface Anal.}\ }\textbf
  {\bibinfo {volume} {24}},\ \bibinfo {pages} {811--820} (\bibinfo {year}
  {1996})}\BibitemShut {NoStop}%
\bibitem [{\citenamefont {Feldman}\ and\ \citenamefont
  {Mayer}(1986)}]{Feldman1986}%
  \BibitemOpen
  \bibfield  {author} {\bibinfo {author} {\bibfnamefont {L.~C.}\ \bibnamefont
  {Feldman}}\ and\ \bibinfo {author} {\bibfnamefont {J.~W.}\ \bibnamefont
  {Mayer}},\ }\href@noop {} {\emph {\bibinfo {title} {Fundamentals of Surface
  and Thin Film Analysis}}}\ (\bibinfo  {publisher} {North Holland-Elsevier},\
  \bibinfo {address} {N.Y.},\ \bibinfo {year} {1986})\BibitemShut {NoStop}%
\bibitem [{\citenamefont {Moya}\ \emph {et~al.}(1993)\citenamefont {Moya},
  \citenamefont {Moya}, \citenamefont {Juv{\'{e}}}, \citenamefont
  {Tr{\'{e}}heux}, \citenamefont {Grattepain},\ and\ \citenamefont
  {Aucouturier}}]{Moya1993}%
  \BibitemOpen
  \bibfield  {author} {\bibinfo {author} {\bibfnamefont {F.}~\bibnamefont
  {Moya}}, \bibinfo {author} {\bibfnamefont {E.~G.}\ \bibnamefont {Moya}},
  \bibinfo {author} {\bibfnamefont {D.}~\bibnamefont {Juv{\'{e}}}}, \bibinfo
  {author} {\bibfnamefont {D.}~\bibnamefont {Tr{\'{e}}heux}}, \bibinfo {author}
  {\bibfnamefont {C.}~\bibnamefont {Grattepain}},\ and\ \bibinfo {author}
  {\bibfnamefont {M.}~\bibnamefont {Aucouturier}},\ }\href
  {https://doi.org/10.1016/0956-716X(93)90439-Y} {\bibfield  {journal}
  {\bibinfo  {journal} {Scr. Metall. Mater.}\ }\textbf {\bibinfo {volume}
  {28}},\ \bibinfo {pages} {343--348} (\bibinfo {year} {1993})}\BibitemShut
  {NoStop}%
\bibitem [{\citenamefont {Gai}, \citenamefont {Smith},\ and\ \citenamefont
  {Owen}(1990)}]{Gai1990}%
  \BibitemOpen
  \bibfield  {author} {\bibinfo {author} {\bibfnamefont {P.~L.}\ \bibnamefont
  {Gai}}, \bibinfo {author} {\bibfnamefont {B.~C.}\ \bibnamefont {Smith}},\
  and\ \bibinfo {author} {\bibfnamefont {G.}~\bibnamefont {Owen}},\ }\href
  {https://doi.org/10.1038/348430a0} {\bibfield  {journal} {\bibinfo  {journal}
  {Nature}\ }\textbf {\bibinfo {volume} {348}},\ \bibinfo {pages} {430--432}
  (\bibinfo {year} {1990})}\BibitemShut {NoStop}%
\bibitem [{sup()}]{supplementary}%
  \BibitemOpen
  \href@noop {} {\ }\bibinfo {note} {See supplementary material at [URL will be
  inserted by AIP Publishing] for details on Cu depth profile and comments on
  adsorption-controlled growth}\BibitemShut {NoStop}%
\end{thebibliography}

%%%%%%%%%%%%%%%%%%%%%%%%%%%%%%%%%%%%%%%%%%%%%% Bibliography %%%%%%%%%%%%%%%%%%%%%%%%%%%%%%%%%%%%%%%%%%%%%%%%%%%%%%%%%%%%

%aipnum4-2.bst 2019-01-14 (MD) hand-edited version of apsrev4-1.bst

%%%%%%%%%%%%%%%%%%%%%%%%%%%%%%%%%%%%%%%%%%%%%%%%%%%%%%%%%%%%%%%%%%%%%%%%%%%%%%%%%%%%%%%%%%%%%%%%%%%%%%%%%%%%

\pagebreak

\listoffigures

\end{document}